\documentstyle[aps,pra,psfig,amstex]{revtex}
\newcommand{\al}{\alpha}
\newcommand{\be}{\beta}

\begin{document}
\preprint{M\'exico ICN-UNAM 99-03}
\title{$H_3^{(2+)}$ molecular ion in a strong magnetic field: 
                       a  triangular configuration} 
\author{ J. C. L\'opez V. ${}^\dagger$}
\address{Instituto de Ciencias Nucleares, UNAM, Apartado Postal 70-543,
        04510 M\'exico D.F., M\'exico}
\date{\today}

\maketitle

\smallskip

\begin{abstract}
  The bound state in the system of three protons and an electron
  $(pppe)$ under a homogeneous strong magnetic field where the protons
  are situated in the vertices of an equilateral triangle
  perpendicular to the magnetic field lines is found. It is shown that
  for magnetic fields $B = 10^{11} - 4.414\times 10^{13}G$ the
  potential energy curves as a function of the internuclear distance
  $R$ have an explicit minimum. For all magnetic fields studied, the
  binding energy of the triangular configuration is less than the
  binding energy of the linear parallel configuration (A.  Turbiner et
  al. {\it JETP Lett. {\bf 69}}, p. 844).  In the contrary to the
  linear case, the binding energy decreases with a magnetic field
  growth, while the equilibrium internuclear distance slowly
  increases.
\end{abstract}


\widetext


\bigskip

Recently a quantitative study of the system of three protons and an
electron $(pppe)$ in a strong magnetic field gave theoretical evidence
of the existence of a bound state for magnetic fields $B>10^{11} G$
\cite{Turbiner:1999}. It manifests the existence of the exotic
molecular ion $H_3^{(2+)}$.  Many years ago it was indicated by
Kadomtsev and Kudryavtsev \cite{Kadomtsev:1971}, and Ruderman
\cite{Ruderman:1971} the possible existence of linear exotic molecular
systems in presence of a strong magnetic field.  Due to the enormous
Lorentz force acting on the electronic cloud, it shrinks in the
direction transversal to the magnetic field lines leading to an
effective quasi-one-dimensionality of the studied systems. This makes
the electron-nuclei attraction more effective to compensate the
Coulombic repulsion of nuclei. Such a fact suggests the possibility of
the existence of exotic systems which do not exist without a magnetic
field. It has been shown some time ago that a strong magnetic field
can lead to the formation of linear hydrogenic chains $H_n, \ n> 2$
situated along magnetic lines
\cite{Ruderman:1971,Salpeter:1992,Salpeter:1995,Salpeter:1997}.  More
recently the first quantitative study about the possible existence of
an exotic bound state in the system of three protons and an electron
$(pppe)$ in a strong magnetic field was carried out
\cite{Turbiner:1999} and it confirmed existence of $H_3^{(2+)}$ ion.

Although the linear parallel configuration (where protons are situated
on a line parallel to the magnetic field line) for the system $(pppe)$
seems to be optimal for very strong magnetic fields, other possible
configurations have to be considered in order to have a better
understanding of the properties of such systems.

In this article we explore the possibility of the existence of the
molecular ion $H_3^{(2+)}$, where the protons are situated in the
vertices of an equilateral triangle perpendicular to a homogeneous
magnetic field. We limit our present investigation to magnetic fields
$B= 10^{11} - 4.414\times 10^{13} G$ for which a non-relativistic
approach is still valid (for a detailed discussion see
\cite{Salpeter:1997} and references therein).  Our study is restricted
to the ground state within the Born-Oppenheimer approximation.


The present calculations are carried out in the framework of a
variational method. Trial functions are chosen in accordance with the
following criterion (see \cite{Turbiner:1980}---\cite{Turbiner:1989}):
(i) the trial function $\Psi_t(x)$ should include all symmetry
properties of the problem in hand; (ii) the trial function for the
ground state should not vanish inside the domain where the problem is
defined; (iii) the potential $V_t(x)=\frac{{\mathbf\nabla}^2
  \Psi_t}{\Psi_t}$, for which the trial function is an exact
eigenfunction, should reproduce the original potential behavior near
singularities as well as its asymptotic behavior. This prescription
has been successfully applied to the study of the $H_2^+$ ion in
strong magnetic fields giving the more accurate results for $B>0$
\cite{Lopez:1997}, as well as to the study of the linear parallel
configuration of the $H_3^{(2+)}$ and $H_4^{(3+)}$ ions in a strong
magnetic fields \cite{Turbiner:1999,Lopez:1999}.

We consider a system of one electron and three identical
infinitely-heavy centers of unit charge situated on the vertices of an
equilateral triangle with internuclear distance $R$ (which is the
length of the triangle side) lying in a plane perpendicular to the
magnetic field of strength $B$ directed along the $z$ axis, ${\vec B}
= (0,0,B)$.

\begin{figure}[thb]
\caption{Triangular configuration for the $(pppe)$ system in a
  magnetic field ${\vec B}$ directed along the $z$ axis.  The protons
  are situated in the $(x,y)$-plane and their positions are marked by
  black circles.}
\label{fig:1}
\begin{center}
\begin{picture}(250,220)
\put(50,0){\psfig{figure=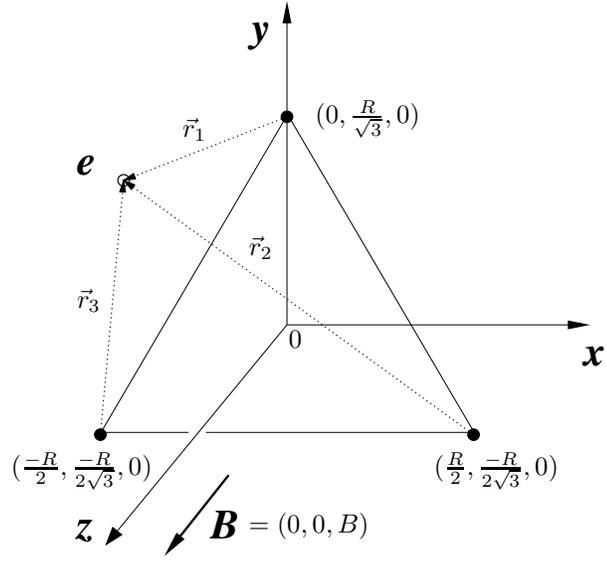,width=200pt,angle=-90} } 
\put(130,80){$0$}
\put(90,160){${\vec r}_1$}
\put(115,115){${\vec r}_2$}
\put(50,95){${\vec r}_3$}
\put(25,32){$(\frac{-R}{2},\frac{-R}{2\sqrt{3}},0)$}
\put(185,32){$(\frac{R}{2},\frac{-R}{2\sqrt{3}},0)$}
\put(140,165){$(0,\frac{R}{\sqrt{3}},0)$}
\put(115,10){$=(0,0,B)$}
\end{picture}
\end{center}
\end{figure}

The potential corresponding to the system we study is given by

\begin{equation}
\label{eq:1}
V=\frac{6}{R} 
-\frac{2}{r_1} - \frac{2}{r_2} - \frac{2}{r_3} +
\frac{B^2 \rho^2}{4}  + B{\hat \ell}_z
\end{equation}
where the quantity $\rho = \sqrt{x^2+y^2}$ is the distance from the
electron position to the $z$-axis, $r_{1,2,3}$ are the distances from
the electron to the first, second and third centers, respectively, $R$
is the distance between centers (see Fig.  ~\ref{fig:1} for notations)
and ${\hat \ell}_z$ is the $z$-component of the angular momentum
operator. The linear Zeeman effect term in the potential $(B\ell_z)$
can be dropped off since we use real trial functions \footnote{For a
  real trial function the expectation value of the $z$-component of
  the angular momentum $\langle \ell_z \rangle$ equals to zero and
  leads to a vanishing contribution to the variational energy emerging
  from the linear Zeeman effect term .} (see below).  Spin Zeeman
effects are neglected.  Through the paper the Rydberg is used as the
energy unit.  For the other quantities standard atomic units are used.


The trial functions we are going to use for the present problem are
similar to those were exploited in \cite{Turbiner:1999} for the study
of the linear parallel configuration.  These functions were built
according to the criterion described in
\cite{Turbiner:1980}---\cite{Turbiner:1989}. They contain the basic
features of Coulomb systems in a magnetic field, as well as the
symmetry property under permutation of the three charged centers. In
particular, it implies that the potentials corresponding to those
functions reproduce the Coulomb-like behavior near the centers and the
two-dimensional oscillator behavior in the $(x,y)$ plane at large
distances (see below).

The simplest of those functions is a function of the Heitler-London
type multiplied by the lowest Landau orbital:

\begin{equation}
  \label{eq:2}
  \Psi_1 = e^{ -\alpha_1(r_1+r_2+r_3) -\beta_1 B \rho^2/4},
\end{equation}
(cf. Eq. (2) in Ref. \cite{Turbiner:1999}), where $\alpha_1$ and $\beta_1$
are variational parameters. Since we consider the distance between
centers as an extra  variational parameter, it has in total three
variational parameters. This function gives an
adequate description of the covalent coupling of the system near
equilibrium. As illustration we show the potential corresponding to
this function:

\begin{eqnarray}
  \label{eq:3}
V_1 =  \frac{{\mathbf\nabla}^2 \Psi_1}{\Psi_1} &=& 
3 \alpha_1^2 -  \beta_1 B 
-2 \alpha_1 \sum_{i=1}^{3} \frac{1}{r_i} + 2 \alpha_1^2
\sum_{i<j}  \, ( {\hat n}_i \cdot {\hat n}_j )
+
\alpha_1 \beta_1 B\sum_{i=1}^{3}  \frac{ \rho^2 -(x x_i + y y_i)}{r_i}
+  \frac{{\beta_1}^2 B^2\rho^2}{4}\ ,
\\
{\hat n}_i \cdot {\hat n}_j &=& \frac{1}{r_i r_j}
\left[
(x-x_i)(x-x_j)+ (y-y_i)(y-y_j)+ (z-z_i)(z-z_j)
\right], \nonumber
\end{eqnarray}
where ${\hat n}_i\ (i=1,2,3)$ is the unit vector in the direction of
the vector pointing from the position of the $i$-th center to the
position of the electron.

A second trial function is a Hund-Mulliken type function multiplied by
the lowest Landau orbital:

\begin{equation}
\label{eq:4}
\Psi_2=
\Big(  {e}^{-\al_2  r_1} + {e}^{-\al_2  r_2} + {e}^{-\al_2  r_3}\Big)
{e}^{-\be_2 B \rho^2 /4} \ ,
\end{equation}
(cf. Eq. (4) in Ref. \cite{Turbiner:1999}). Here $\al_2,\be_2$ are
variational parameters.  This function describes a ionic coupling of a
hydrogen atom with two charged centers.

Another suitable trial function which describes a ionic coupling
between an $H_2^+$ ion and a proton is given by

\begin{equation}
\label{eq:5}
\Psi_3=
\Big( 
 {e}^{-\al_3  (r_1+r_2)} +
 {e}^{-\al_3  (r_1+r_3)} +
 {e}^{-\al_3  (r_2+r_3)} 
\Big)
{e}^{-\be_3 B \rho^2 /4} \ ,
\end{equation}
(cf. Eq. (5) in Ref. \cite{Turbiner:1999}) where $\al_3,\be_3$ are
variational parameters.  Since the $H_2^+$ ion has the lowest total
energy among the one-electron systems for $B \lesssim 10^{13} G$
\cite{Lopez:1999}, an important contribution coming from this trial
function is expected.

In order to include in a single trial function the different physical
behavior near equilibrium and at large distances, appropriate
interpolations of the trial functions (\ref{eq:2}), (\ref{eq:4}) and
(\ref{eq:5}) are done:

\begin{itemize}

\item[(i)] A natural interpolation is given by a non-linear
superposition of the form:
\begin{equation}
    \label{eq:6}
    \Psi_{4-nls}=
    \left(
      \sum_{\{ \al_4, \al_5, \al_6 \} }
     \hspace{-15pt}
     {e}^{ -\al_4 \, r_1 - \al_5 \, r_2 - \al_6 \, r_{3}  }
    \right) {e}^{-\be_4  B {\rho^2}/{4}    },
\end{equation}
(cf. Eq. (6) in Ref. \cite{Turbiner:1999}) where $\al_4, \al_5, \al_6 $
and $\be_4$ are variational parameters, and the sum is over all
permutations of the parameters $\{ \al_4, \al_5, \al_6 \}$.

If all parameters coincide ($\al_4= \al_5 = \al_6 = \al_{1}$), the
function (\ref{eq:6}) reduces to the Heitler-London type function
(\ref{eq:2}). Function (\ref{eq:6}) also reduces to the Hund-Mulliken
type wave function (\ref{eq:4}) when only one parameter is non-zero
say $\al_4 =\al_2$ and $\al_5=\al_6=0$.  If two parameters are
non-zero, and equal, say, $\al_4 = \al_5 = \al_3$, and $\al_6=0$ it
reduces to the trial function (\ref{eq:5}).  When all parameters are
different among them and different from zero the function (\ref{eq:6})
provides us with a $3$-center modification of a Guillemin-Zener type
function.  The function (\ref{eq:6}) has in total five variational
parameters (including the internuclear distance $R$ as one of the
parameters).

\item[(ii)] Another more immediate interpolation is given by a linear
superposition of  the functions (\ref{eq:2}), (\ref{eq:4}) and
(\ref{eq:5}). 

\begin{equation}
\label{eq:7}
\Psi_{5-{ls}}= A_1 \Psi_1 + A_2 \Psi_2 + A_3 \Psi_3 ,
\end{equation}
where $A_1, A_2, A_3$ are taken as extra variational parameters. Since
$\Psi_{1,2,3}$ are not orthogonal, the parameters $A_{1,2,3}$ do not
have the usual meaning of weight factors.

\end{itemize}

For the present calculations we use a linear superposition of the
above interpolations  (\ref{eq:6}), (\ref{eq:7}) given by:

\begin{equation}
\label{eq:8}
\Psi_{6}= A_1 \Psi_1 + A_2 \Psi_2 + A_3 \Psi_3 + A_4 \Psi_{4-nls},
\end{equation}
where $A_1, A_2, A_3, A_4$ are considered again as extra variational
parameters with no meaning of weight factors. The function
(\ref{eq:8}) combines in a single function form, a suitable
description of the system for the different physical regimes (near
equilibrium and large internuclear distances for strong and very
strong magnetic fields). Therefore we guess it should provide a
relevant approximation for the ground state of the system. This was
indeed the case of the linear parallel configuration of $H_3^{(2+)}$
(see Ref.  \cite{Turbiner:1999}). This function has in total fourteen
variational parameters.

The variational procedure is carried out using the standard
minimization package MINUIT from CERN-LIB.  We use the integration
routine D01FCF from NAG-LIB.  All integrals are calculated with
relative accuracy $\gtrsim 10^{-8}$.


The results of our calculations are presented in Table ~I.  For all
considered magnetic fields $ B = 10^{11} - 4.414\times 10^{13} G$ we
found a minimum in the potential energy curve as a function of the
internuclear distance $R$ (see below Fig. 3) indicating the formation
of a bound state. For such magnetic fields the total energy of the
triangular configuration is always larger than the total energy of the
corresponding linear parallel configuration. Indeed this shows that
for very strong magnetic fields the linear configuration is the most
favored. However, a new striking unexpected phenomenon appears when
one considers a triangular configuration.  While for the linear
configuration the binding energy\protect\footnote{The binding energy
  is defined as the affinity to keep the electron bound $E_b = B-E_T$.
  $B$ is given in $Ry$ and thus has a meaning of the energy of free
  electron in magnetic field.} increases and the internuclear distance
decreases as the magnetic field grows, the opposite is true for the
triangular configuration. Namely, the binding energy is slowly
decreasing and the internuclear distance is slowly increasing (!) both
being almost constant.  A similar behavior occurs with the
longitudinal localization length of the electron $ \langle | z |
\rangle$ which also slowly increases in the triangular configuration
with a magnetic field growth, opposite to what happens in the linear
configuration where it decreases rather sharply.  The transversal size
of the electronic cloud $\langle \rho \rangle$ is very close to the
corresponding cyclotron radius $B^{-1/2} (a.u.)$ for both cases. We do
not have a clear physical picture of this phenomenon and it needs
further consideration.

A surprising feature is revealed if we study different trial
functions. In Table II it is shown the variational results for the
total energy $E_T$ for two quite arbitrarily chosen values of the
internuclear distance, $R=0.45$ and $R=2.527$. For magnetic fields
$10^{11} \lesssim B \lesssim 10^{12} \, G$ an adequate description of
the system for internuclear distances $R \simeq 0.45$ is given by the
trial function (\ref{eq:5}) modelling a ionic coupling, $H_2^+ \, +
$proton. In this case, the single function (\ref{eq:5}) gives the
lowest total energy among the trial functions $\Psi_{1,2,3,4}$.  On
the other hand, the trial function (\ref{eq:2}) modelling a covalent
coupling  gives the lowest total
energy for $R\simeq 2.527$ giving a hint  that a covalent
coupling dominates  when the internuclear distance is large
compared to the equilibrium distance.  Amazingly, for $B\gtrsim 10^{13}
G$ the variational energy seems to be insensitive to the specific form
of the trial function (!).

The electronic density distribution $|\Psi|^2$ in the plane $z=0$ is
shown in Fig. 2. For all magnetic fields $B=10^{11} - 4.414\times
10^{13} \,G$ the electronic density distribution $|\Psi|^2$ exhibits a
single maximum located at the center of the triangle $(x=0,y=0)$. As
the magnetic field grows the electronic cloud becomes more
concentrated around $x=0,y=0$, while its distance to the protons is
slowly increased.


From our results we can draw the conclusion that the exotic molecular
ion $H_3^{(2+)}$ can exist in the presence of a strong magnetic field
in a triangular configuration of the charged centers which is unstable
towards a decay mode to the linear configuration.  In particular, in
the present article we have shown that the molecular ion $H_3^{(2+)}$
can also exist in the equilateral (fixed nuclei) triangular
configuration perpendicular to the magnetic field.

The results of our variational study for the ground state show that,
contrary to the case of the linear parallel configuration, the binding
energy of the system decreases, and the internuclear distance
increases, as the magnetic field grows from $B=10^{11} G$ up to
$4.414\times 10^{13}G$.  A more extended study of this as well as
other configurations is needed in order to have a better understanding
of the properties of the exotic system $H_3^{(2+)}$. In particular, a
proper contribution coming from the linear Zeeman effect term in
(\ref{eq:1}) has to be taken into account. However, the fact that the
linear Zeeman effect contribution vanishes in our consideration (which
occurs when real trial functions are considered) looks relevant in the
region of very strong magnetic fields $B=10^{11} - 4.414\times 10^{13}
\,G$ where this term can be neglected anyhow.

Although in general it is not clear so far (see discussion in
\cite{Salpeter:1997}) how an adiabatic separation of the electronic
and the nuclear motion can be performed in the presence of a strong
external magnetic field, our fixed-nuclei approach is a good starting
point to study of the exotic molecular ion $H_3^{(2+)}$.

\acknowledgments

The author wish to express his deep gratitude to Alexander Turbiner
for introducing to the subject and for numerous valuable discussions.

This work is supported in part by DGAPA Grant \# IN105296 (M\'exico).



\newpage

\begin{table}[htb]
\label{table1}
\caption{Variational results for the triangular configuration of the
  $H_3^{(2+)}$ ion in a strong magnetic field. The total energy $E_T$ and the  
  binding energy $E_b=B-E_T$, are in $Ry$. 
  The equilibrium internuclear distance  ${R_{eq}}$ and the expectation
  values of the longitudinal and transversal localization length of
  the electron $\langle | z | \rangle$ and $\langle | \rho |
  \rangle$   are in $a.u.$. The 
  corresponding quantities  for the
  $H_3^{(2+)}$ ion in the linear parallel configuration  were taken from
  \protect\cite{Turbiner:1999}. The energy of free electron and its
  cyclotron radius $ B^{-1/2}$ are also shown.}
\begin{tabular}{cccccc}
\( B \) (Gauss)&
\( E_{T} \)&
\( E_{b} \)&
\( R_{eq} \)&
\( <|z|> \)&
\( <\rho > \)\\
\hline 
\hline 
\multicolumn{6}{c}{Triangular Configuration}\\
\hline 
\( 10^{11} \)&
40.5591&
1.985&
0.6237&
0.561&
0.192\\
\( 10^{12} \)&
423.722&
1.719&
0.6414&
0.595&
0.061\\
\( 10^{13} \)&
4252.724&
1.690&
0.6645&
0.610&
0.019\\
\( 4.414\times 10^{13} \)&
18777.31&
1.674&
0.6757&
0.622&
0.009\\
\hline 
\multicolumn{6}{c}{Linear Configuration}\\
\hline 
\( 10^{11} \)&
36.429&
6.1151&
0.803&
0.864&
0.186
\\
\( 10^{12} \)&
410.296&
15.144&
0.346&
0.438&
0.060 
\\
\( 10^{13} \)&
4220.09&
34.324&
0.165&
0.242&
0.019
\\
\( 4.414\times 10^{13} \)&
18723.88&
55.103&
0.110&
0.168&
0.009 
\\
\hline 
\multicolumn{6}{c}{Free electron}\\
\hline 
\( 10^{11} \)&
42.5441
&
&
&
&
0.153
\\
\( 10^{12} \)&
425.441
&
&
&
&
0.048 
\\
\( 10^{13} \)&
4254.41
&
&
&
&
0.015
\\
\( 4.414\times 10^{13} \)&
18778.98
&
&
&
&
0.007 
\\
\end{tabular}
\end{table}


\begin{table}
\label{table2}
\caption{Comparison of the total energy $E_T$ for the triangular
  configuration of the system $(pppe)$ for two arbitrary 
  internuclear distances $R=0.45$ and $R=2.527$ obtained with
  trial functions $\Psi_{1,2,3,4}$, and with their linear superposition
  given by trial function $\Psi_6$.} 
\begin{tabular}{ccccccc}
\( B \) (Gauss)&
\( R \)&
\( \Psi _{1} \)&
\( \Psi _{2} \)&
\( \Psi _{3} \)&
\( \Psi _{4} \)&
\( \Psi _{6} \)\\
\hline 
&
0.45&
41.10121 &
41.12298 &
41.09077 &
41.09591 &
41.08986 \\
\( 10^{11} \)&
&
&
&
&
&
\\
&
 2.527&
41.58129 &
41.58405 &
41.58146 &
41.58131 &
41.58122 \\
\hline 
&
0.45&
423.9111 &
423.9195 &
423.9115 &
 423.9111 &
423.9110 \\
\( 10^{12} \)&
&
&
&
&
&
\\
&
 2.527&
424.4875&
424.4877 &
424.4876 &
424.4876&
424.4865 \\
\hline 
&
0.45&
4252.956 &
4252.956 &
4252.956 &
4252.956 &
4252.956\\
\( 10^{13} \)&
&
&
&
&
&
\\
&
 2.527&
4253.461&
4253.461&
4253.461&
4253.461&
4253.460\\
\hline 
&
0.45&
18777.53 &
18777.53 &
 18777.53 &
18777.53 &
18777.53\\
\( 4.414\times 10^{13} \)&
&
&
&
&
&
\\
&
 2.527&
 18778.03&
18778.03&
18778.03&
18778.03&
18778.03\\
\end{tabular}
\end{table}

\pagebreak

\begin{figure}[tb]
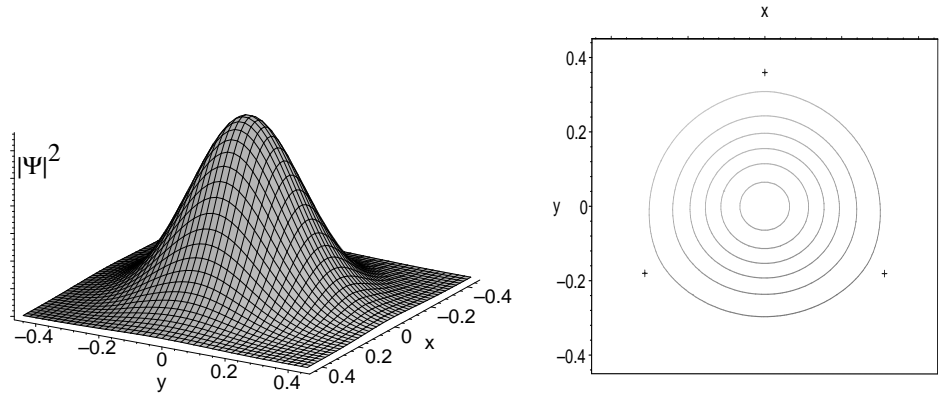

\caption{Electronic probability density $|\Psi|^2$ for  the $H_3^{(2+)}$ 
  molecular ion in the triangular configuration at $z=0$ and their
  corresponding contours for (a) $B=10^{11} G$, (b) $B=10^{12} G$, (c)
  $B=10^{13} G$. The position of the charged centers is indicated by
  crosses.  Normalization of $|\Psi|^2$ is not fixed.}
\label{Fig:2}
\[
 \begin{array}{ccc}
 \psfig{figure=b100A.ps,width=2.2in,angle=-90}   & &
 \psfig{figure=b100B.ps,width=2.2in,angle=-90} \\[-30pt]
& \hspace{20pt} (a) \hspace{20pt}  & \\
 \psfig{figure=b1000A.ps,width=2.2in,angle=-90}  & &
 \psfig{figure=b1000B.ps,width=2.2in,angle=-90} \\[-30pt] 
& \hspace{20pt} (b) \hspace{20pt} & \\
 \psfig{figure=b10000A.ps,width=2.2in,angle=-90} & &
 \psfig{figure=b10000B.ps,width=2.2in,angle=-90} \\[-30pt]
&\hspace{20pt}  (c) \hspace{20pt} & \\
 \end{array}
 \]
\end{figure}

\begin{figure}[tb]
\caption{Potential energy curves as function of the internuclear
  distance $R$ for the $H_3^{(2+)}$ molecular ion in the triangular
  configuration for magnetic field:  (a) $B=10^{11} G$, (b)
  $B=10^{12} G$, (c) $B=10^{13} G$. The top of the energy scale
  corresponds to the value of the energy of free electron $E_e =
  B$.}
\label{Fig:2}
\[
\begin{array}{c}
\psfig{figure=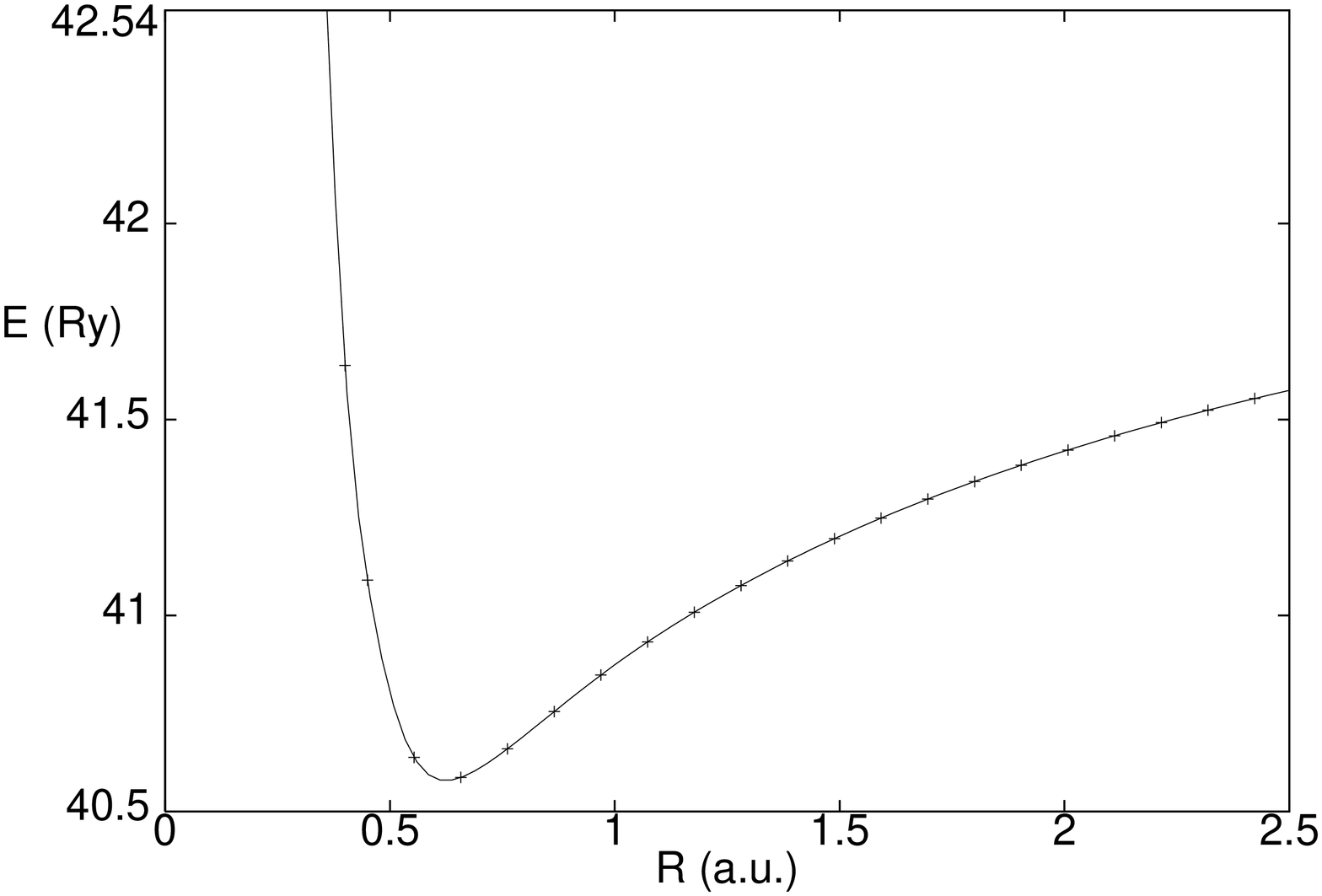,width=3.25in,angle=-90} \\
(a) \\
\psfig{figure=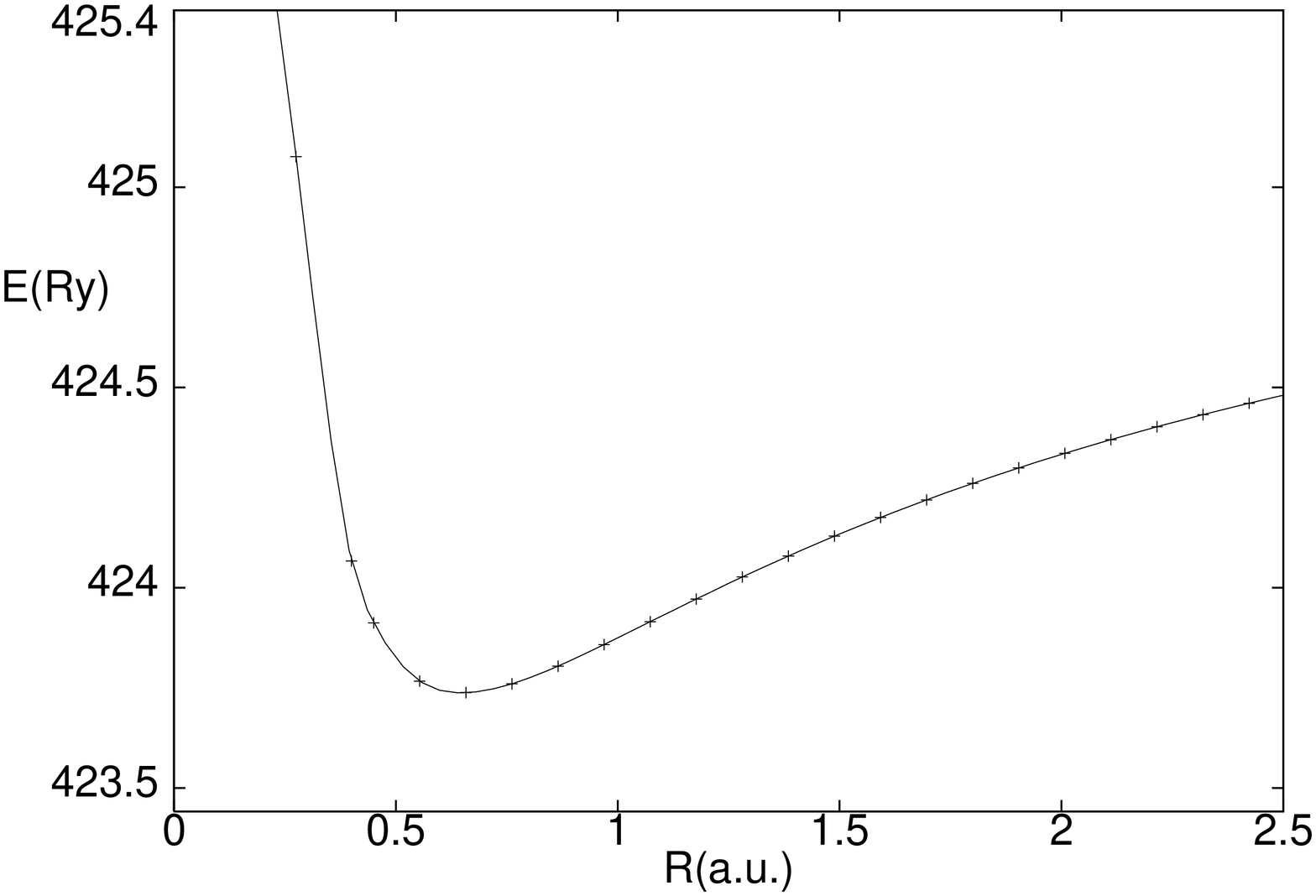,width=3.25in,angle=-90} \\
(b) \\
\psfig{figure=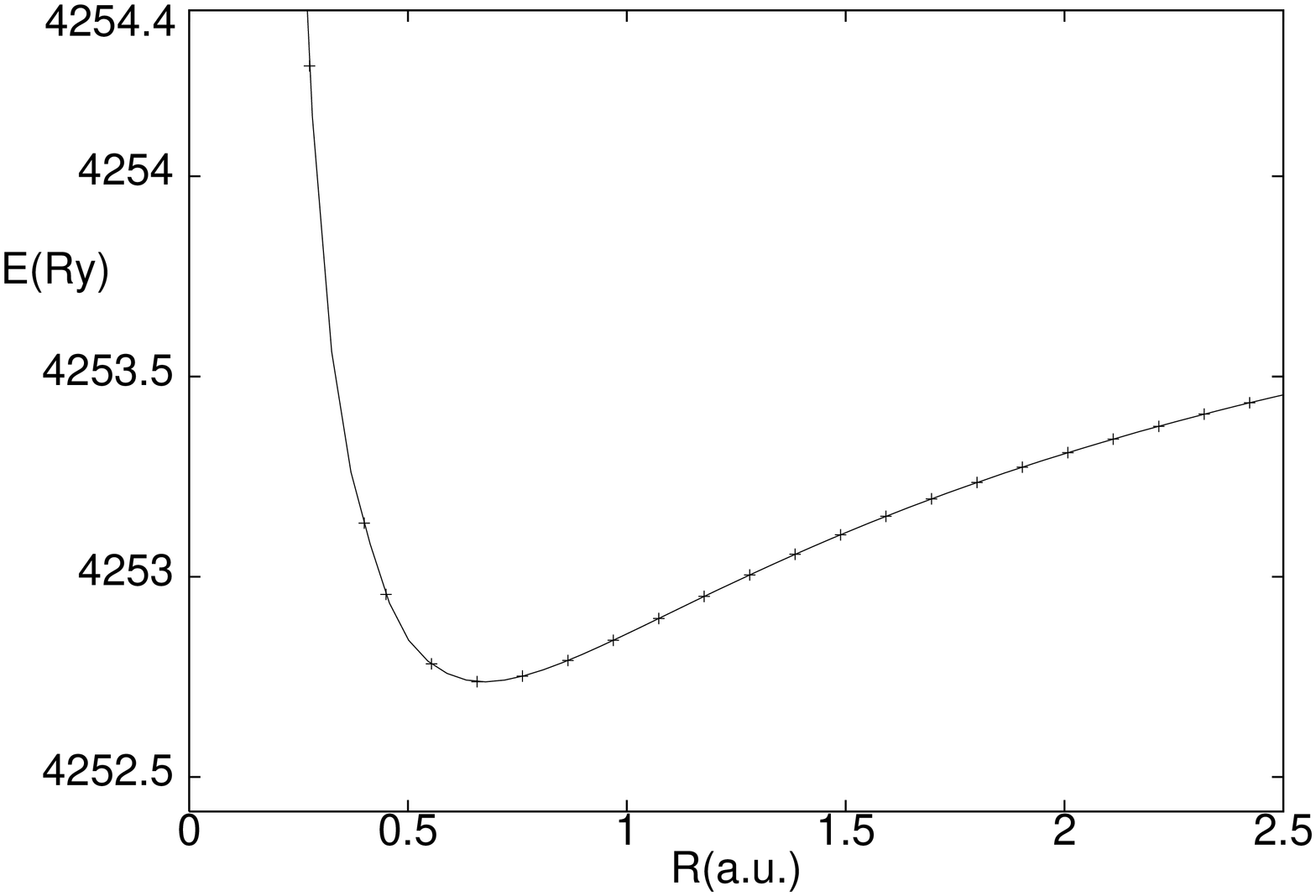,width=3.25in,angle=-90} \\
(c)
\end{array}
\]
\end{figure}


\end{document}